\documentstyle[editedbook,epsfig,psfig,epsf]{mq}

\def\cm2{cm$^2$ }
\def\se1{s$^{-1}$ }

\begin{opening}
\title{Pair Annihilation and Radio Emission from Nova Muscae}
\author{D.C. Hannikainen$^{1,2}$ \& C.R. Kaiser$^2$}
\institute{$^1$ Observatory, PO Box 14, 00014 University of Helsinki, Finland. \\
$^2$ Department of Physics and Astronomy, University of Southampton, Southampton, UK.}
\end{opening}

\runningtitle{Pair annihilation from Nova Muscae}
\runningauthor{Hannikainen \& Kaiser}

\begin{document}
\vspace{-0.5cm}
\begin{abstract}
{\small In the hard X-ray spectra of some X-ray binaries line features 
around 500~keV are detected.
We interpret these as arising from pair annihilation in relativistic 
outflows leading to a
significant Doppler shift of the frequencies of the lines. 
We show that a small fraction of pairs escaping the 
annihilation region may give rise to the radio 
synchrotron emission observed in Nova Muscae 1991.}
\end{abstract}

\section{Introduction}
The composition of the jets observed in Galactic X-ray
transients, either electron-positron or
electron-proton plasma, is still not fully established.  

Here we review the arguments presented in \cite{Kaiser}
 where we proposed that the annihilation line features observed
 in the hard X-ray spectra of some X-ray transients arise from pairs in a 
 bipolar outflow. 
At the time of annihilation, this outflow is
 already accelerated to relativistic bulk speeds causing a significant
 Doppler shift of the frequency of the annihilation lines. 
We also showed
 that the subsequent emission of radio synchrotron radiation from the
 outflow may be caused by only a small fraction of pairs escaping from
 the annihilation region. 
We apply this idea to radio and hard X-ray observations of Nova Muscae 1991.
The energy requirements for
 this source rule out a large contribution of protons to the outflow.

\section{Doppler-shifted annihilation lines}


{\em Conditions for a strong, narrow annihilation line.}
The direct annihilation of an electron-positron pair results in the
 production of two $\gamma$-ray photons, each with an energy of
 511\,keV in the rest-frame of the annihilating particles. 
In non-thermal plasmas relativistic electrons or pairs may be injected
 into the plasma. 
The injected pairs and those produced in the plasma
 may cool to sub-relativistic energies and thermalize before
 annihilating, thus leading to a narrow annihilation line \cite{Lightman}. 
If the line is strong,
 it can rise above the Comptonization spectrum and becomes
 detectable. 
This requires a high pair yield, $Y$, defined as the ratio
 of the energy converted to pairs and the energy supplied to the
 plasma. 
The highest pair yields can be achieved when the plasma is
 `photon-starved', i.e. when the number of injected relativistic
 photons strongly exceeds that of the injected soft photons 
 \cite{Zdziarski}. 
In this case, $Y\sim 0.25$ and a
 strong, narrow annihilation line above the Comptonization continuum
 becomes observable.

The observation of a narrow annihilation line most 
 likely indicates a plasma with strong injection of non-thermal
 electrons or pairs. 
The injection of leptons into a spherical volume
 of radius $R$ is characterised by the compactness (e.g. \cite{Lightman})
 $l_{\rm e} = L_{\rm e} \sigma _{\rm T} / R m_{\rm e} c^3$,
 where $\sigma _{\rm T}$ is the Thomson cross-section and
 $L_{\rm e}$ is the power of the electron injection,
$L_{\rm e} = (4 \pi R^3 / 3) m_{\rm e} c^2 \int Q_{\rm l}(\gamma)
\left( \gamma -1 \right) \, d\gamma$.
Here, $Q_{\rm l}(\gamma)$ is the rate of injection of
 leptons with Lorentz factor $\gamma$ per unit volume per unit time per
 unit $\gamma$. 
If the compactness $l_{\rm e}$ can be inferred from the
observations of an annihilation line, then 
 the above equations
 can be used to constrain $Q_{\rm l}(\gamma)$. \\


\noindent{\em Relativistic Doppler-shifts.}
Any emission of relativistically moving material is Doppler-shifted in
 its frequency. 
For material moving with bulk velocity $v_{\rm b}=\beta
 c$ at an angle $\theta$ to the line of sight to the observer, the
 observed frequency, $\nu$, of radiation emitted at frequency $\nu'$ in
 the rest-frame of the material is given as

\begin{equation}
\nu = \frac{\nu'}{\gamma _{\rm b} \left( 1 \pm \beta \cos \theta
\right)} = \nu' \delta_{\pm},
\label{delta}
\end{equation}

\noindent where $\gamma _{\rm b}$ is the Lorentz factor corresponding
 to the velocity $\beta$ (e.g. \cite{Rybicki}). 
The upper signs correspond to material receding
 along the line of sight to the observer while the lower signs indicate
 approaching material. 
From Equation (\ref{delta}) it is
 clear that radiation of material receding from an observer is always
 redshifted. 
However, for approaching material the emission may be
 blueshifted or redshifted, depending on the combination of $\beta$ and
 $\theta$. 
Solving Equation (\ref{delta}) for $\beta$ we find
that for $\beta > 2 \cos \theta / \left( 1 + \cos ^2 \theta \right)$ the
annihilation line arising from the approaching jet material will be
redshifted.
The range of velocities which result in such a
 Doppler redshift is largest for angles to the line of sight close to
 $90^{\circ}$.



\section{The case of Nova Muscae}

\begin{figure}[htb]
\centering
\psfig{file=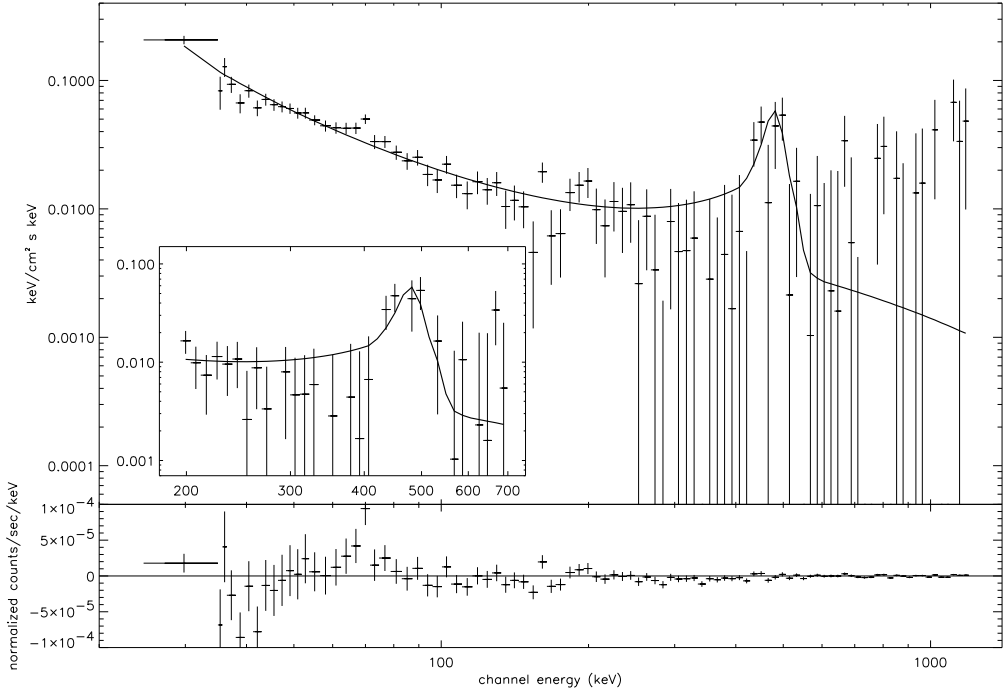,width=14cm}
\caption{The best-fitting model to the spectrum of Jan~21. The large figure shows the full
spectrum from the last third of the Jan~20--21 observation, showing the emergence of the
annihilation features. The inset shows the data points around 474~keV used in the fitting, while the bottom panel shows the residuals of the model to the data in normalized
counts/second.
}
\label{fig:fig2}
\end{figure}

Nova Muscae (GRS 1124$-$684) was discovered on 1991~Jan~8 by
 Granat (\cite{Lund}, \cite{Sunyaev91})
 and Ginga \cite{Makino}. 
During monitoring observations, a
 strong, narrow line near 500 keV was detected on Jan~20--21 which 
 had not been noted before. 
The line flux was observed to increase during the
 observation within the space of a few hours. 
Unfortunately, the
 observations stopped before the line flux decreased again, implying a
 lifetime of the line emission of at least 10 hours. 
Simultaneously to
 the strong line near 500 keV, there was also increased emission
 detected near 200 keV. The spectrum obtained with Granat during
 1991~Jan~20--21 is shown in Figure~\ref{fig:fig2}.

\noindent The original discovery also triggered a radio monitoring programme
 using the Molongolo Observatory Synthesis Telescope (MOST) 
 at 843\,MHz (\cite{Kesteven}, \cite{Ball}).
\cite{Ball} note four distinct features in the lightcurve. 
There is a general decay of the radio flux from the very beginning of the
 programme continuing until about January~22. 
There is another flare observed from January~31 until around February~5. 
Finally, a short flare lasting only one day was
 observed by MOST at 843\,MHz on January~24 with a measured flux
 density of 24\,mJy. 
In the following we will concentrate on the detection of the
 $\gamma$-ray lines on January~21 and the brief radio flare on
 January~24.  
We speculate that the same ejection event is responsible
 for the emission at opposite ends of the electromagnetic spectrum
 observed on the two days. \\


\noindent{\em Constraints from the annihilation line.}
We assume here that the two $\gamma$-ray lines observed in the
 spectrum of Nova Muscae arise from pair annihilation in a relativistic
 bipolar outflow from the very centre of the system. 
In this model, the
 line at 474\,keV is associated with the approaching component while
 that at 194\,keV arises from the receding component.
We find $\theta = 60^{\circ}\pm7^{\circ}$ for the angle to the line of sight 
 of the component motion and $\beta = 0.84 \pm 0.02$ for the bulk velocity of
 the components. 
This corresponds to a Lorentz factor of $\gamma _{\rm b} = 1.86\pm0.12$. \\

The rate of annihilation implied by the line flux 
 ($(6.6\pm3.6) \times 10^{-3}$\,photons s$^{-1}$ cm$^{-2}$) 
 is $\dot{N}_+\sim 2\times 10^{43}$\,pairs s$^{-1}$ for a distance of 5.5\,kpc
 \cite{Orosz}.  
This enormous rate is sustained for at least
 10 hours \cite{Sunyaev92}. 
It is therefore very
 unlikely that this feature at 474~keV arises from a large number
 of pairs formed practically instantaneously and then slowly
 annihilating away. 
This would imply a short-lived
 flash of annihilation photons with a fast, exponential decay
 contrary to the lifetime of the annihilation features of at least
 10 hours. 
A more promising approach is to assume that the pair
 producing plasma is in equilibrium, i.e. the annihilation losses
 are balanced by pair creation.
We fit the spectrum using the model NTEEA within the XSPEC package which is an
 implementation of the model developed in \cite{Lightman}.
The only free parameters in this fit were the compactness of the non-thermal
 electron injection, $l_{\rm e}$, and the compactness of injected soft
 photons, $l_{\rm s}$.
Figure~\ref{fig:fig2} shows our best fit with a reduced $\chi^2$-value 
 of 1.35 (for 84~d.o.f). 
The free parameters,
 $l_{\rm e} \sim 3000$ and $l_{\rm s} \sim 50$, imply a strongly 
 photon-starved plasma.
However, the values of the model parameters are not
 well-constrained and reasonable fits to the data can be 
 obtained for $l_{\rm e} > 100$.
Re-arranging the equation above yields
$L_{\rm e} = R m_{\rm e} c^3 l_{\rm e} / \sigma _{\rm T}$.
Substituting in our lower limits for $R$ and $l_{\rm e}$, we
 find $L_{\rm e} \ge 7 \times 10^{30}$\,W. 
It follows that $Q_0 = 9\times
 10^{26}$\,m$^{-3}$ s$^{-1}$ for this lower limit. 
The rate at which
 relativistic electrons are injected is then $\dot{N}_{\rm inj}=3\times
 10^{43}$\,particles s$^{-1}$. 
This is comparable to the observed
 annihilation rate of $2\times10^{43}$\,particles s$^{-1}$. \\


\noindent{\em Implications of a bipolar flow.}
In the case of an
 ejection event shorter than 10 hours, the annihilation rate would decrease
 dramatically as the ejected material travels outwards and expands.
In the model presented here
 we argue that the bulk of the outflow containing the pairs is already
 accelerated when they annihilate, thus explaining the redshifts of the
 two observed lines. 
The velocity of the outflow is then 0.84\,c.

The energy required to drive the outflows is enormous. 
As the relativistic electrons necessary for pair production are highly
 relativistic, their mass in the rest frame of the outflow material is
 given by $\gamma m_{\rm e}$. 
Therefore, the total kinetic power of the
 relativistic electrons, as measured in the source rest frame, injected
 into the outflow can be approximated as 
 $\dot{E}_{\rm kin} = V Q_0 m_{\rm e} c^2 \left( \gamma _{\rm b} -1
 \right) \int \gamma ^{1-p} \, d\gamma \sim 6\times 10^{31} {\rm \ W}$.
In this estimate we have used the limiting values for the
 electron compactness, $l_{\rm e} = 100$, and the size of the emission
 region, $R = 2\times 10^5$\,m. 
Any increase in these parameters will also cause the energy estimate to rise. 
Therefore, the lower limit on 
 $\dot{E}_{\rm kin}$ corresponds already to about 80\% of the Eddington
 luminosity of a 6\,M$_{\odot}$ black hole. 
Balance of electrical
 charge requires the presence of positively charged particles in the
 outflow. 
If these are protons with negligible thermal
 energy but travelling at the necessary bulk velocity, then another
 $4\times 10^{33}$\,W in kinetic energy are required. 
This is at least a factor 50 more than the Eddington luminosity. 
This energy injection
 into the outflow has to be sustained for more than 10 hours and thus
 makes a large proton content in the jet very unlikely.

The only alternative is then that from the very beginning of the
 ejection the outflow material in the annihilation region consists of
 virtually a pure pair plasma.  
This removes the requirement of
 relatively inefficient pair production from relativistic
 electrons. 
The pairs must be injected into the relativistic bulk flow
 with a relativistic velocity distribution to explain the strength of
 the annihilation line \cite{Maciolek}, the cooling within the outflow and then
 annihilation at a rate of $2\times 10^{43}$\,s$^{-1}$. 
In this case,
 the required kinetic power is about $6\times 10^{30}$\,W, or 8\% of
 the Eddington luminosity.



\section*{Acknowledgments}
We thank Marat Gil'fanov for providing us with the original {\it Granat}
data.

\end{document}